\documentclass[aps,prl,amsfonts,amsmath,twocolumn,superscriptaddress]{revtex4-1}
\usepackage{amsmath}
\usepackage{amssymb}
\usepackage{multirow}
\usepackage{graphicx}
\usepackage{grffile}
\newcommand{\bse}{\begin{subequations}}
\newcommand{\ese}{\end{subequations}}
\newcommand{\be}{\begin{equation}}
\newcommand{\ee}{\end{equation}}
\newcommand{\bea}{\begin{eqnarray}}
\newcommand{\eea}{\end{eqnarray}}
\newcommand{\al}{\alpha}
\newcommand{\ba}{\beta}

\newcommand{\cU}{\mathcal U}

\begin{document}

%\preprint{}

\title{Universality versus material dependence of fluctuation forces
  between metallic wires}


%\thanks{}

\author{E. Noruzifar}
\affiliation{Department of Physics and Astronomy,
  University of California, Riverside, California 92521, USA}

\author{T. Emig}
\affiliation{Laboratoire de Physique Th\'eorique et Mod\`eles
 Statistiques, CNRS UMR 8626, Universit\'e Paris-Sud, 91405 Orsay,
 France}

\author{R. Zandi}
\affiliation{Department of Physics and Astronomy,
  University of California, Riverside, California 92521, USA}
\affiliation{Laboratoire de Physique Th\'eorique et Mod\`eles
 Statistiques, CNRS UMR 8626, Universit\'e Paris-Sud, 91405 Orsay,
 France}


\date{\today}

\begin{abstract}
  We calculate the Casimir interaction between two parallel wires and
  between a wire and a metall plate.  The dielectric properties of the
  objects are described by the plasma, Drude and perfect metal models.
  We find that at asymptotically large separation interactions
  involving plasma wires and/or plates are independent of the material
  properties, but depend on the dc conductivity $\sigma$ for Drude
  wires. Counterintuitively, at intermediate separations the
  interaction involving Drude wires can become independent of
  $\sigma$.  At smaller separations, we compute the interaction
  numerically and observe an approach to the proximity approximation.
\end{abstract}

%\pacs{}
%\keywords{}

\maketitle

%\section{}
%\subsection{}
%\subsubsection{}

Effective interactions between nanowires and nanotubes have attracted
lots of attention due to their growing applications in micro- and
nanomechanical systems \cite{fen2003, Cra2000,husain03}. The knowledge
of the interactions between single walled carbon nanotubes (SWCNT)
with different chirality and hence electromagnetic response is
important to separate a polydisperse solution of SWCNT in fractions of
equal chirality \cite{podgo}. Under many circumstances, van der Waals or
Casimir forces are the dominant interaction and hence a precise
understanding of them is needed.  Furthermore, cylindrical shapes are
important for precision Casimir force measurements, in comparison to
spheres, because of the larger effective area of interaction
\cite{brown05, decca10}.  Approximations of the Casimir
force between cylinders and plates \cite{decca10} have shown that the
temperature dependence varies based on the description
of the material properties.  Thus there is a need for exact
calculations of the Casimir force for cylindrical shapes taking into
account realistic material response.

%[more applications in the context of biophysics, capillary phenomena, 
%see first paragraph of Barash and Kyasov. Should we mention them here
%as well?]

It has been demonstrated that Casimir interactions strongly depend on
the combined effects of shape and material properties, see, e.g.,
\cite{umar, zandi10}. The interplay is particularly strong for quasi
one-dimensional conducting materials due to strongly anisotropic
collective charge fluctuations. Indeed, for two parallel perfectly
conducting wires of distance $d$ the retarded interaction energy per length
is ${\cal E}/L \sim \hbar c/d^2$, apart from a logarithmic factor
\cite{emig06}. It decays only slowly compared to the retarded
interaction ${\cal E}/L \sim \hbar c R^4/d^6$ between insulating cylinders
that do not support collective fluctuations. Most studies of
interactions between one-dimensional systems over a wide range of
separations concentrate on these two situations. However, low
dimensionality in combination with finite conductivity and plasmon
excitations should give rise to interesting new effects that might be
probed experimentally using, e.g., the coupling to mechanical
oscillation modes. The often employed technique for these effects, the
proximity force approximation (PFA) cannot capture the correlations of
shape and material response since it is based on the interaction
between planar surfaces. There have been attempts to compute the van
der Waals interaction between cylinders (and plates) for particular
frequency dependent permittivities
\cite{barash89,dobson06,dalvit06,emig06,rahi08,drummond07,dobson09}.
However, the interplay between shape and material effects is not
transparent in these works as they are limited either to perfect
metals or to asymptotic limits.

In this Letter, we employ the scattering approach to study the
interaction between two infinitely long, parallel metallic wires and a
wire and a plate that are described either by the plasma or the Drude
dielectric function. We model the wires as circular cylinders of
radius $R$ and obtain analytical results for the interactions at
distances much larger than $R$. We find regimes of different amplitude
and power-law, depending on the relation of the
radius to the length $\lambda_\sigma=2\pi c/\sigma$ with
conductivity $\sigma$ and the plasma wave length $\lambda_p$.  

Most interestingly, we find that the interaction involving Drude wires
approaches the universal interaction between perfect metal wires at
intermediate separations and becomes {\it non-universal} (material
dependent) at distances $d \gtrsim R^2/\lambda_\sigma$. This behavior
is explained in terms of the size of collective charge fluctuations in
a Drude metal. For wires that support plasma oscillations, the
interaction does show universality at asymptotically large distances.
In order to confirm the validity of our large distance expansions and
to compare with the PFA at short distances we have performed numerical
computations of the interaction.

To investigate the Casimir interaction between a cylinder parallel to
a plate or another cylinder, we employ the scattering approach, which
allows us to calculate the energy over a large range of separations
between the objects. Because of the translational symmetry along the
cylinder axis, the Casimir interaction can be written as \cite{rahi09}
\begin{equation}
\label{energy_general}
{\mathcal E} = \frac{\hbar c L}{4 \pi^2} \int_0^{\infty} 
d \kappa \int_{-\infty}^{\infty} dk_z\, \ln\det({\mathbf 1}-{\mathbb N})\,,
\end{equation}
with $L$ the length of the cylinder, $\kappa$ the Wick rotated
frequency, $k_z$ the wave number along the cylinder axis, and $\bf 1$
the identity matrix.  The matrix $\mathbb N$ factorizes into the
object's scattering amplitudes (T-matrices) that encode the material
dependence and distance dependent translation matrices that described
the coupling between multipoles on distinct objects.

{\it Parallel wires} -- For two infinitely
long parallel cylinders with radii $R$, aligned along the $z$-axis,
the elements of the matrix $\mathbb N$ for polarizations $\alpha$,
$\beta=E,\, M$ and partial waves $m$, $m'$
are
\begin{equation}
  \label{eq:N_2_cyl}
  {\mathbb N}_{mm'}^{\alpha\beta} = \sum_{\gamma=E, M}T_{m}^{\alpha\gamma}
\sum_{n=-\infty}^\infty {\cU_{mn}^{12}}\, T_{n}^{\gamma\beta}\, {\cU_{nm'}^{21}} \,, 
\end{equation}
with $T$ the T-matrix of the cylinder (see below).  The translation
matrix, $\cU^{12}$, relates the regulare vector waves ${\mathbf
  M}_{k_z n}^\text{reg}=(\kappa^2+k_z^2)^{-1/2}\nabla\times
[I_n(p\rho)e^{i(n\theta + k_z z)}]$ and ${\mathbf N}_{k_z
  n}^\text{reg}=\kappa^{-1} \nabla\times {\mathbf M}_{k_z
  n}^\text{reg}$ with $p=\sqrt{\kappa^2+k_z^2}$ in the cylindrical
coordinates $(\rho,\theta,z)$ of one cylinder to the outgoing vector
waves of the other cylinder that are given through
replacing the Bessel functions $I_n$ by the Bessel functions of second
kind, $K_n$. The elements are given by
\be
  \label{eq:B-matrix-WR}
 {{\mathcal U}^{12}_{n n'}}=(-1)^{n'}\, K_{n-n'} \left(p\, d\right) \, \quad
 {\mathcal U}^{21}_{n n'} = (-1)^{n-n'} {\mathcal U}^{12}_{n n'} 
\,,
\ee
where $d$ is the distance between the centers of cylinders.  Note that
the translation matrix conserves polarization and thus is independent
of the polarization index.  The T-matrix for a cylinder with
dielectric function $\epsilon(ic\kappa)$ and magnetic permeability $\mu(ic\kappa)$ 
is diagonal in the partial wave number $n$ but couples
polarizations. Its elements are
\begin{align}
   \label{eq:T-matrix-elements-ee}
  T^{EE}_{n} &= -\frac{I_n(pR)}{K_n(pR)} 
\frac{\Delta_{2,n}\Delta_{3,n} +P_n^2}{\Delta_{1,n}\Delta_{2,n}+P_n^2}\,,\\
  \label{eq:T-matrix-elements-me}
  T^{EM}_{n} &= - \frac{P_n}{ \sqrt{\epsilon\mu} (pR)^2 K_n^2(pR)}
\frac{1}{\Delta_{1,n}\Delta_{2,n} +P_n^2} \,,
\end{align}
 with $P_n = ({n k_z}/{(\sqrt{\epsilon\mu} R^2 \kappa)}) 
\left( {1}/{p'^2} -{1}/{p^2}\right)$, $p'=\sqrt{\epsilon\mu\kappa^2+k_z^2}$
and 
\begin{equation}
  \label{Delta}
  \Delta_{1,n} =  \frac{I'_n(p'R)}{p' R I_n(p'R)} -\frac{1}{\epsilon} \frac{K'_n(pR)}{pR K_n(pR)}\,.
\end{equation}
$\Delta_{2,n}$ is obtained from Eq.~(\ref{Delta}) by interchanging
$\epsilon$ with $\mu$, and $\Delta_{3,n}$ follows from
Eq.~(\ref{Delta}) by replacing $K'_n$ with $I'_n$ and $K_n$ with
$I_n$. The elements $T^{MM}_{n}$ are given by $T^{EE}_{n}$ after
interchanging $\epsilon$ with $\mu$. Finally, antisymmetry in
polarization yields $T^{ME}_{n}=-T^{EM}_{n}$.

In order to investigate the impact of the material properties on the
Casimir interaction, we consider plasma, Drude and perfect metal
cylinders with the magnetic permeability $\mu= 1$.  The Drude 
dielectric function is
\be
\label{di-fun}
\epsilon (i c \kappa) = 
1+ \frac{(2\pi)^2}{(\lambda_{\rm p}\kappa)^2+\lambda_{\sigma}\kappa/2}\,,
\ee
and reproduces the plasma model for $\lambda_{\sigma}\to 0$.  Using
Eq.~(\ref{di-fun}), the asymptotic behavior of the T-matrix element of
Eq.~\eqref{eq:T-matrix-elements-ee} for $n=0$ at small frequencies
($\kappa \ll 1$, $k_z/\kappa$ fixed) is given by
\be
\label{tmatrix_cyl}
T_0^{EE}  \approx -\frac{p^2R^2}{
C(\kappa) - p^2R^2 \ln(p R/2)}\,,
\ee
with $C(\kappa) \approx {\lambda_{\rm p}}^2 \kappa^2\, /(2\pi^2 )$ if
$R\ll \lambda_p$, $C(\kappa)={\lambda_\sigma \kappa}/{(4\pi^2)}$ if
$\kappa \ll \lambda_\sigma/\lambda_p^2$, $1/\lambda_\sigma$, and
$C(\kappa)=0$ for plasma, Drude and perfect metal cylinders,
respectively. At small frequencies $\kappa$ but fixed $k_z/\kappa$,
one has for Drude cylinders $T^{EE}_0 \sim \kappa$, while for plasma
and perfect metal cylinders one has $T^{EE}_0 \sim 1$.  Since
$T^{MM}_0\sim\kappa^2$, $T^{EM}_0=T^{ME}_0=0$ and higher order
elements associated with $n\ne 0$ scale as $\kappa^{2|n|}$, it is
sufficient for the large distance interaction to consider only the element
$T_0^{EE}$.

\begin{figure}[ht]
\centering
\includegraphics[width=1.\linewidth]{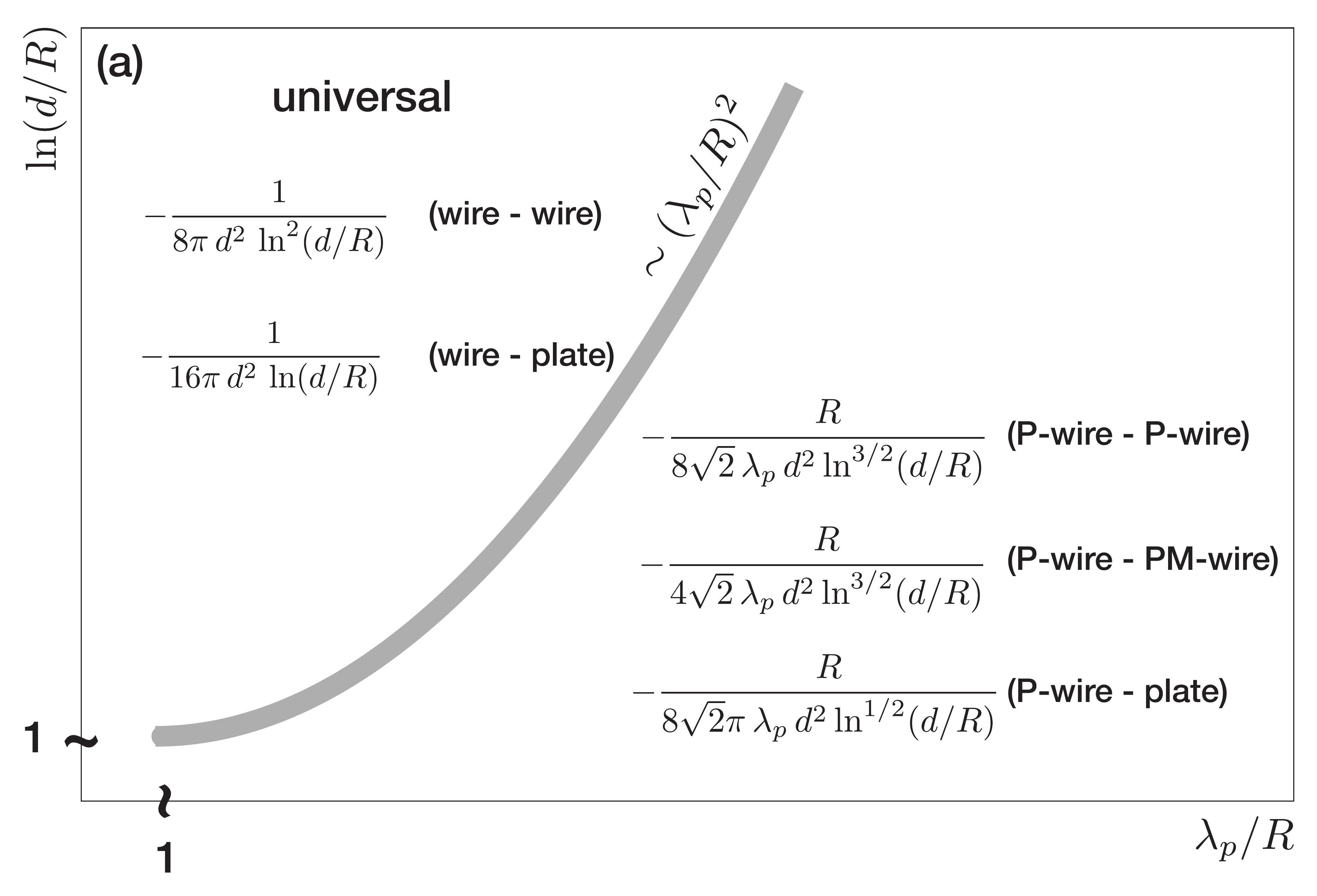}
\includegraphics[width=1.\linewidth]{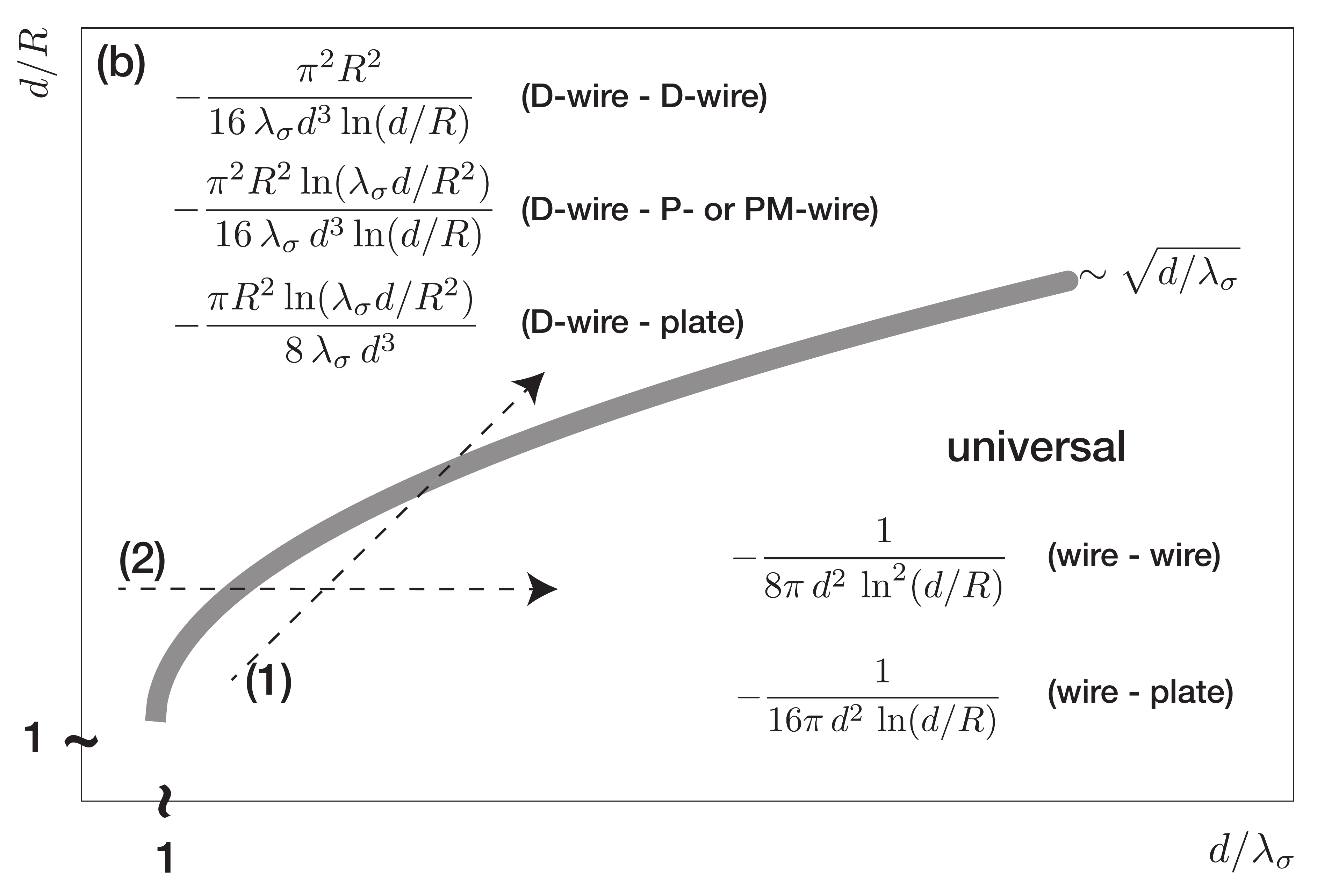}
\caption{\label{crossovers} Summary of the different forms of
  interaction between two wires and a wire and a plate. Shown are the
  rescaled interaction energies per cylinder length, ${\cal E}/(\hbar c L)$. (a)
  Interaction involving a plasma (P) wire with another
  plasma wire, a perfect metal (PM) wire or a plate. The asymptotic
  results apply sufficiently far away from the separating
  curve $\ln(d/R)\sim (\lambda_p/R)^2$ and for $d/R$, $\lambda_p/R \gg
  1$. (b) Interaction involving a Drude (D) wire with
  another Drude wire, a plasma wire, a perfect metal wire or a plate.
  The separating curve is given, up to logarithmic corrections, by
  $d/R \sim \sqrt{d/\lambda_\sigma}$. The shown expressions hold for
  $d/R$, $d/\lambda_\sigma \gg 1$ and $d \gg
  \lambda_p^2/\lambda_\sigma$. Depending on the relative size of length scales,
different regimes can be reached: Dashed arrow (1)
  corresponds to an increasing distance $d$ which ultimately leads to
  a {\it non-universal} interaction. Dashed arrow (2) indicates an
  overall increase of the geometry (i.e., $d/R$ fixed) with constant
  conductivity leading to a {\it universal} interaction.}
\end{figure}

The Casimir interaction between metallic wires is in general
complicated and no simple analytical expression that applies to all
length scales can be obtained. However, using
Eqs.~(\ref{energy_general}) and (\ref{eq:N_2_cyl}) along with
$T_0^{EE}$ given in Eq. (\ref{tmatrix_cyl}), the asymptotic
interaction at large separations, $d \gg R$, can be calculated in
various limiting cases. For metals with diverging response at zero
frequency, one usually observes universal behavior for the interaction
at large separations. Indeed, that is what we obtain for the
interaction between plasma wires (or plasma and perfect metal wires)
which then becomes ${\cal E}=-{\hbar c L }/{(8\pi d^{2}\ln^2 (d/R))}$
\cite{emig06, rahi08}. This universal form is only applicable beyond
an exponentially large crossover length $d\sim R \exp(\lambda_{\rm
  p}^2/R^2)$.  Below this scale the interaction becomes material
dependent. For two plasma wires or a plasma wire and a perfect metal
wire, with $\lambda_{\rm p}/R\gg 1$, the energy scales as
$-R/(\lambda_p d^2 \ln^{3/2}(d/R))$. For numerical coefficients, see
Fig.~\ref{crossovers}(a). 

For configurations involving at least one Drude cylinder we find a
rather distinct behavior that deviates from naive expectations for
universality. For large distances $d \gg R, \, \lambda_\sigma$ we
obtain two different scaling regimes that are separated, up to
logarithmic corrections, by the curve $d/R \sim
\sqrt{d/\lambda_\sigma}$, see Fig.~\ref{crossovers}(b). The unexpected
feature is that the interaction is universal in the regime where $d\ll
R^2/\lambda_\sigma$. If the distance is increased beyond this
crossover scale (with all other length scales kept fixed, see arrow
(1) in Fig.~\ref{crossovers}(b)), the interaction becomes material
dependent and up to logarithmic corrections, scales as
$-R^2/(\lambda_\sigma d^3)$ for a Drude wire interacting with another
Drude wire or a plasma or perfect metal wire.  For detailed forms of
the interactions in this limit, see Fig.~\ref{crossovers}(b).  If,
however, the radii of the wires are increased in the same way as their
distance ($d/R$ fixed, see arrow (2) in Fig.~\ref{crossovers}(b)),
finite conductivity becomes unimportant at large distances and the
interaction assumes the perfect metal form.  An intuitive explanation
of this non-universal large distance behavior is given below. Note
that the decay of interactions between insulating wires for $d\gg R$
scales as $\hbar c L R^4/d^6$ with a material dependent coefficient.

{\it Wire parallel to a plane} -- Now we consider a wire that is
parallel to a plate.  We assume that the plate is in the $y-z$ plane
and its distance from the cylinder ($z$) axis is $d$.  Then the
matrix $\mathbb{N}$ in Eq.~\eqref{energy_general} can be written as
\begin{equation}
\begin{split}
\raisetag{50pt}
\label{Mmatrix}
\mathbb{N}_{m m'}^{\al\ba} &=
 \sum_{\gamma,\gamma'=E,M} T_{m}^{\al\gamma}\int_{-\infty}^{\infty} dk_y 
\frac{e^{-2d\sqrt{{\bf k}_{\bot}^2+ \kappa^2}}}{2\sqrt{{\bf k}_{\bot}^2+\kappa^2}} \\
&\times D_{m k_z \gamma,{\bf k}_{\bot}\gamma'}
\;T^{\gamma'}_{{\bf k}_{\bot}} \;
 D_{{\bf k}_{\bot}\gamma',m' k_z \beta}^{\dagger}\;(1-2\delta_{\gamma',\beta}) \,, 
\end{split}
\end{equation}
where $k_y$ is $y$-component of the wave vector, ${\bf k}_{\bot}
\equiv (k_y, k_z)$, and the matrix ${D}_{m k_z \gamma,{\bf
    k}_{\bot}\gamma'}$ converts between plane and cylindrical waves
\cite{rahi09}.  Further, $T^\gamma_{{\bf k}_{\bot}}$ are the
$T$-matrix elements for the plate which are given by the usual Fresnel
coefficients \cite{rahi09}. For perfect metal plates $T^{E}_{{\bf
    k}_{\bot}}=T^{M}_{{\bf k}_{\bot}} = 1$ and for small $\kappa$ at
fixed $k_\perp/\kappa$ one has for the plasma model $T^{E}_{{\bf
    k}_{\bot}}=T^{M}_{{\bf k}_{\bot}} = 1+{\cal O}(\lambda_p \kappa)$
and for the Drude model $T^{E}_{{\bf k}_{\bot}}=T^{M}_{{\bf k}_{\bot}}
= 1+{\cal O}(\lambda_\sigma \kappa)$.  Due to this
identical behavior of the plate's T-matrix at small $\kappa$, the
interaction at asymptotically large distances is independent of the
model that describes the metal plate. 

We calculate the Casimir energy between a wire and a plate at large
separations $(d/R\gg1)$, using Eqs.~(\ref{energy_general}),
(\ref{tmatrix_cyl}) and (\ref{Mmatrix}).  We again find two different
scaling regimes that are separated by curves that are given by the
same expressions that we found for two wires, see
Fig.~\ref{crossovers}. In the  universal regime (perfect metal), the interaction
is ${\cal E}=-{\hbar c L }/{(16\pi d^{2}\ln^2 (d/R))}$ \cite{emig06,
  rahi08} for a plasma wire  at asymptotically large $d\gg R$ and for a
Drude wire at intermediate distances with $\lambda_\sigma, \,
\lambda_p^2/\lambda_\sigma\ll d\ll R^2/\lambda_\sigma$, see
Fig.~\ref{crossovers}(b). In the other regime the interaction is
non-universal with the energies up to logarithmic accuracy scaling as
${\cal E}\sim \hbar c L R/(\lambda_p d^2)$ for a plasma wire and
${\cal E}\sim \hbar c L R^2/(\lambda_\sigma d^3)$ for a Drude wire.
For the precise form see Fig.~\ref{crossovers}(b).  This can be
compared to the faster decay that is observed for an insulating wire
interacting with a plane which has ${\cal E}\sim \hbar c LR^2/d^4$
\cite{rahi09}.

\begin{figure}[ht]
\centering
\includegraphics[width=1\linewidth]{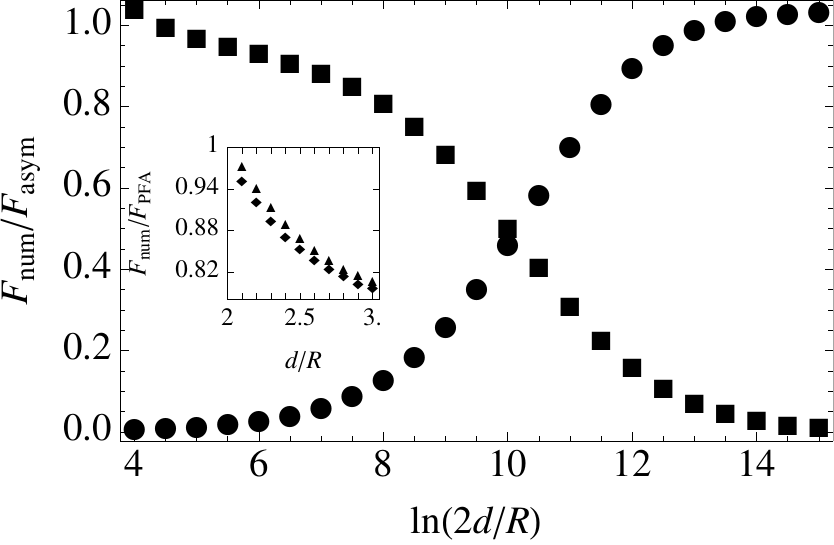}
\caption{\label{pfa} Ratio of the numerically computed force and the
  universal (squares) and non-universal (circles) asymptotic results
  of Fig.~\ref{crossovers} for two Drude wires with $\lambda_{\rm
    p}/R=0.5$ and $\lambda_{\sigma}=\lambda_{\rm p}/27.4$.  The inset
  shows the ratio of the numerical force to the proximity force
  approximation (PFA) for Drude cylinders with $\lambda_{\rm p}/R=0.5$
  (diamonds) and $\lambda_{\rm p}/R=0.05$ (triangles).  }
\end{figure}

{\it Numerical results} - It is interesting to compare the asymptotic
results with exact numerical calculations to identify the regions in
which the asymptotics are correct.  We calculate numerically the
integrals and determinant in Eq.~(\ref{energy_general}).  The infinite
matrix $\mathbb N$ given by Eq.~(\ref{eq:N_2_cyl}) is truncated at a
finite number $n=n_{\rm max}$ of partial waves.  $n_{\rm max}$ depends
on the separation between the objects and is chosen such that the
energy varies by less than $0.01\%$ upon increasing $n_{\rm max}$ by
$10$. As the separation becomes shorter, $n_{\rm max}$
increases. While for $d/R = 10$, $n_{\rm max} = 9$ is sufficient, we
need to use $n_{\rm max}=91$ for $d/R=2.1$.

Figure \ref{pfa} shows the ratio of the numerically computed force
between two Drude wires and the corresponding asymptotic results
(universal and non-universal regimes, see Fig.~\ref{crossovers}(b))
versus $\ln(2d/R)$. The material parameters are chosen as
$\lambda_{\rm p}/R=0.5$ and $\lambda_{\sigma}=\lambda_{\rm p}/27.4$
which correspond to gold with $\lambda_{\rm p}=137$ nm and
$\lambda_{\sigma}\approx 5$ nm. At intermediate separations, the force
normalized to the universal result approaches unity whereas at
asymptotically large separations the force normalized to the
non-universal result tends to unity.  This confirms the validity of
the crossover shown in Fig.~\ref{crossovers}(b).  We also compare our
numerical results for the force at short separations with those
obtained from the proximity force approximation (PFA) which is based
on the Lifshitz formula for parallel plates made of the same material
as the wires.  The inset in Fig.~(\ref{pfa}) shows the ratio of our
numerical result for the force and the PFA result versus $d/R$ for
Drude wires with $\lambda_{\rm p}/R=0.5$ and $\lambda_{\rm p}/R=0.05$.
Our data support the validity of the PFA in the limit of vanishing separations.

In summary, we have calculated the Casimir force between two metallic
cylinders and a metallic cylinder parallel to a plate. We find that a
significant feature of the interaction between a Drude Cylinder with
another Drude cylinder or a plate is that upon increasing the
separation, the interaction can move from a universal regime to a 
non-universal one.  This behavior can be understood from the wave
equation for the electric field inside a Drude cylinder.  For
imaginary frequencies $\omega=ic\kappa$, the Helmholtz operator
$\nabla^2 + \epsilon(\omega) (\omega/c)^2$ for a good Drude conductor
becomes $\nabla^2 - 8\pi^2 \kappa/\lambda_{\sigma}$. We are interested
in the maximal scale of the field and hence charge fluctuations for a
given $\kappa$. With the smallest transverse wave vector $k_x$,
$k_y\sim 2\pi/R$ we find 
the dispersion relation
\begin{equation}
  \label{eq:dispersion}
  |k_z| \sim R^{-1}\sqrt{\kappa/\kappa_c -1}\, , \quad \kappa_c =
  \lambda_\sigma/R^2 \, .
\end{equation}
Hence, collective charge fluctuations on arbitrarily large scales
exist only for $\kappa > \kappa_c$ which is a consequence of
dimensionality that does not appear in the absence of transverse
constraints ($R\to\infty$).  For $\kappa<\kappa_c$ charge fluctuations
break up into clusters of typical size $\sim
R/\sqrt{1-\kappa/\kappa_c}$ due to finite conductivity. The spectral
contribution to the interaction between cylinders at distance $d$ is
peaked around $\kappa\sim 1/d$. If $d \lesssim 1/\kappa_c$
($d/R\lesssim \sqrt{d/\lambda_\sigma}$, see Fig.~\ref{crossovers}(b)),
collective charge fluctuations contribute strongly to the interaction
and render it universal as for perfect metal cylinders for which
$\kappa_c\sim 1/\sigma \to 0$. In the asymptotic regime with $d
\gtrsim 1/\kappa_c$ ($d/R\gtrsim \sqrt{d/\lambda_\sigma}$, see
Fig.~\ref{crossovers}(b)), finite conductivity prevents fluctuations on
arbitrarily large scales and hence the interaction is proportional to
$\sigma$, i.e., non-universal.  It is important to note that as $R$
goes to zero, $\kappa_c$ becomes larger, and in consequence the finite
conductivity of cylinder becomes more important.

Finally, an estimate of the interaction between two gold wires based on the
asymptotic expressions in Fig.~\ref{crossovers} with $R=10$nm, length
$L=100\mu$m, $\lambda_p=137$nm and $\lambda_\sigma=5$nm at a distance
$d=200$nm yields a force of $\approx 1$pN within the plasma description
and $\approx 27$pN within the Drude model. These forces are
experimentally detectable and allow a much clearer distinction between
plasma and Drude model predictions as compared to two plates or a plate and
sphere \cite{zandi10}. This is of particular importance in view of
recent experimental findings that interactions between
metals might not be consistent with the Drude model \cite{Decca07}. 

\acknowledgements We thank M.~Kardar and U.~Mohideen for useful
conversations regarding this work.  This work was supported by the
NSF through grants DMR-06-45668 (RZ), DARPA contract
No.~S-000354 (RZ and TE).

\bibliography{refs.bib}

\end{document}